\documentclass[twoside,12pt]{article}
\usepackage{CJK}
\usepackage{indentfirst}
\usepackage{bm}
\usepackage{epsfig}
\usepackage{amssymb}
\usepackage{amsmath}

\begin{document}


\thispagestyle{empty} \vspace*{0.8cm}\hbox
to\textwidth{\vbox{\hfill\huge\sf Commun. Theor. Phys.\hfill}}
\par\noindent\rule[3mm]{\textwidth}{0.2pt}\hspace*{-\textwidth}\noindent
\rule[2.5mm]{\textwidth}{0.2pt}


\begin{center}
\LARGE\bf Unusual slow energy relaxation induced by mobile discrete breathers in one-dimensional lattices with next-nearest-neighbor coupling$^{*}$
\end{center}

\footnotetext{\hspace*{-.45cm}\footnotesize $^*$DX was supported by the start-up fund of Minjiang university and NSF (Grant No.
2021J02051) of Fujian Province of China. The work of CX was supported by the start-up fund of Minjiang university. The work of WZ was supported by the NNSF (Grant No. 12105133) of China and NSF (Grant No. 2021J011030) of Fujian Province of China.}
\footnotetext{\hspace*{-.45cm}\footnotesize $^\dag$Corresponding author, E-mail: t10077@fzu.edu.cn }
\footnotetext{\hspace*{-.45cm}\footnotesize $^\ddag$Corresponding author, E-mail: xmuxdx@163.com }

\begin{center}
\rm Bin Xu$^{\rm a)}$, \ \ Jun Zhang$^{\rm a)\dagger}$, \ \ Wei Zhong$^{\rm b)}$, \ \ Chi Xiong$^{\rm b)}$, \ and  \ Daxing Xiong$^{\rm b)\ddagger}$
\end{center}

\begin{center}
\begin{footnotesize} \sl
${}^{\rm a)}$ Department of Physics, Fuzhou University, Fuzhou 350108, Fujian, China \\
${}^{\rm b)}$ MinJiang Collaborative Center for Theoretical Physics, College of Physics and Electronic Information Engineering, Minjiang University, Fuzhou 350108, China \\
\end{footnotesize}
\end{center}

\begin{center}
\footnotesize (Received XXXX; revised manuscript received XXXX)

\end{center}

\vspace*{2mm}

\begin{center}
\begin{minipage}{15.5cm}
\parindent 20pt\footnotesize
We study the energy relaxation process in one-dimensional (1D) lattices with next-nearest-neighbor (NNN) couplings. This relaxation is produced by adding damping (absorbing conditions) to the boundary (free-end) of the lattice. Compared to the 1D lattices with on-site potentials, the properties of discrete breathers (DBs) that are spatially localized intrinsic modes are quite unusual with the NNN couplings included, i.e., these DBs are mobile, and thus they can interact with both the phonons and the boundaries of the lattice. For the interparticle interactions of harmonic and Fermi-Pasta-Ulam-Tsingou-$\beta$ (FPUT-$\beta$) types, we find two crossovers of relaxation in general, i.e., a first crossover from the stretched-exponential to the regular exponential relaxation occurring in a short timescale, and a further crossover from the exponential to the power-law relaxation taking place in a long timescale. The first and second relaxations are universal, but the final power-law relaxation is strongly influenced by the properties of DBs, e.g. the scattering processes of DBs with phonons and boundaries in the FPUT-$\beta$ type systems make the power-law decay relatively faster than that in the counterparts of the harmonic type systems under the same coupling. Our results present new information and insights for understanding the slow energy relaxation in cooling the lattices.
\end{minipage}
\end{center}

\begin{center}
\begin{minipage}{15.5cm}
\begin{minipage}[t]{2.3cm}{\bf Keywords:}\end{minipage}
\begin{minipage}[t]{13.1cm}
Energy relaxation; Discrete breathers; Next-nearest-neighbor coupling
\end{minipage}\par\vglue8pt

\end{minipage}
\end{center}

\section{Introduction}
Discrete breathers (DBs), also known as intrinsic localized modes, are spatially localized nonlinear vibrational modes in defect-free discrete systems. In the past decades, huge numbers of theoretical and experimental studies had been devoted to confirm the existence and stability of DBs. Most of the theoretical studies focused on the anharmonic chains at zero temperature, from which the existence of DBs has been proved exactly (see refs.\cite{DB-1,DB-2} and reviews\cite{DB-3,DB-4}). Since then, in other periodic systems such as cantelever arrays\cite{DB-5}, Josephson junction arrays\cite{DB-6,DB-7}, electrical lattices\cite{DB-8,DB-9}, mass-spring chains\cite{DB-10}, arrays of coupled pendula\cite{DB-11}, chains of magnetic pendulums\cite{DB-12}, and granular crystals\cite{DB-13,DB-14}, the existence of DBs has been proved experimentally. There are also many studies in real crystals\cite{DB-R}, for example, in ionic NaI\cite{DB-15}, covalent Si, Ge, and diamond\cite{DB-16,DB-17},  A-uranium\cite{DB-18},  ordered alloys\cite{DB-19} as well as proteins\cite{DB-20,DB-21}.

Standing DBs are localized modes of system energy. However, sometimes DBs can move with a small velocity. These mobile excitations can thus interact with the boundaries. It is thus interesting to study how this kind of DBs affects the energy relaxation at a finite-temperature system when the boundaries play a role. By using the cooling method, i.e., imposing the absorbing conditions on the boundaries, Aubry and Tsiron\cite{Re-1} first observed the stretched-exponential lattice energy relaxation induced by standing DBs in one-dimensional (1D) lattices with on-site potentials, in contrast to the standard exponential relaxation law of the corresponding linear chains. Subsequently, the counterpart two-dimensional lattices have been considered and a similar conclusion has been drawn in \cite{Re-2}. Further detailed studies\cite{Re-3,Re-4} of the harmonic and anharmonic chains without on-site potentials showed that the stretched-exponential energy relaxation is not universal, instead a usual exponential relaxation law is observed in a short time, followed by a power-law decay over a long time. The mechanism of the change of the power-law relaxation is induced by standing DBs. As the interactions of standing DBs with the boundaries are very weak, the energy release is actually suppressed and it is thus difficult to observe equilibrium states on typical simulation time scales. As a result, the lattice is in a metastable state, called the nonequilibrium residual state, similar to the glassy patterns in disordered systems.

In addition to the above progress, it would be worth noting that the properties of DBs have been found to be related to other physical properties. For example, by considering some complex crystal structures, such as the next-nearest-neighbor (NNN) interactions\cite{Xiong-1,Xiong-2}, the lattice with altering mass or interactions\cite{Xiong-3}, the on-site potentials \cite{Onsite-1,Onsite-2,Onsite-3} and the long-rang interactions\cite{LR}, DBs have been studied in the perspective of energy transport. In such cases, DBs are more evident and sometimes can be moveable. This causes that this kind of DBs can be the effective scatters of both phonons and boundaries\cite{Scatter}, hence they can reduce the thermal conductivity\cite{Onsite-3,LR}. We also note that even only considering the role of nonlinearity, DBs can already affect some other macroscopic properties\cite{Others-1}, e.g., elastic constants\cite{Others-2}, thermal expansion\cite{Others-3,Others-4}, and heat capacity\cite{Others-2,Others-4,Others-5}.

The purpose of this paper is to study the slow energy relaxation in a more complicated situation including mobile DBs. To this end, we shall focus on the Fermi-Pasta-Ulam-Tsingou-$\beta$ (FPUT-$\beta$) chain with NNN interactions. By properly adjusting the ratio of the NNN coupling to the nearest-neighbor (NN) coupling, one might be able to produce the mobile DBs\cite{Xiong-1,Xiong-2}. Therefore, by employing such models we are devoted to exploring the combined effects of NNN interactions and nonlinearity (which produces moveable DBs) on the slow energy relaxation. The rest of this article is organized as follows: In Sec. 2 we describe the reference models; Sec. 3 presents the cooling methods to study energy relaxation and the physical quantities of interest; Sec. 4 provides the main results for different lattice models, from which the role of NNN interactions and nonlinearity will be demonstrated; Finally, Sec. 5 draws our conclusion.

\section{Models}
\begin{figure}[bt]
\centerline{\epsfig{file=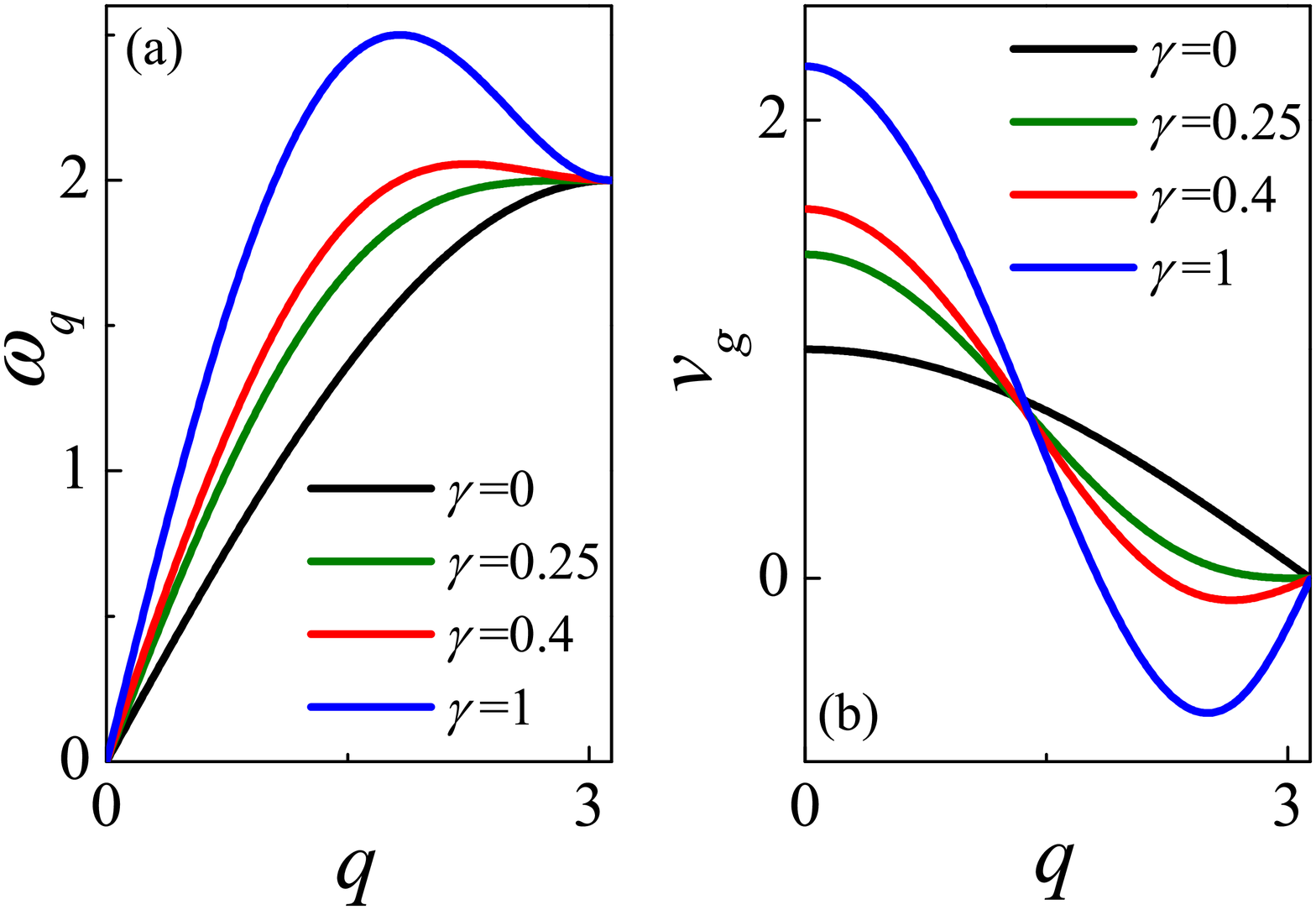,width=12cm,height=8cm,clip=}}
\vspace*{8pt}
\caption{(a) Phonon frequency $\omega_q$ and (b) group velocity $v_g$ versus wave number $q$ for 1D lattice without on-site potentials. From bottom to top, the curves correspond to $\gamma = 0$, $0.25$, $0.4$, and $1$, respectively.}
\label{fig1}
\end{figure}

A general 1D lattice with both NN and NNN interparticle interactions and without on-site potential can be represented by
\begin{equation}
H=\sum_{i}^N \left[\frac{p_{i}^{2}}{2}+ V(x_{i+1}-x_i) +\gamma V(x_{i+2}-x_i)  \right],
\label{HH}
\end{equation}
where $p_i$ is the $i$th (totally $N$ particles and all with unit mass) particle's momentum, $x_i$ is its displacement from equilibrium position, $V(\xi)$ is the interparticle potential and the parameter $\gamma$ specifies the comparative strength of the NNN to NN couplings. We use the FPUT-$\beta$ interparticle interaction $V(\xi)=\frac{1}{2}\xi^2+ \frac{1}{4}\xi^4$ to mainly consider the combined role of NNN coupling and nonlinearity in energy relaxation. For such a model, it would be worth noting that it has a special phonon dispersion relation (see Fig.~\ref{fig1} for the phonon spectrum): $\omega_q=2 \sqrt{ \sin^2 (q/2) + \gamma \sin^2 q}$, where $q$ is the wave number for phonons and $\omega_q$ is the corresponding frequency. From this dispersion relation, one obtains the phonon's group velocity $v_g=\mathrm{d} \omega_q / \mathrm{d} q = \omega^{-1} [\sin q+2 \gamma \sin(2q)]$ and finds that: (i) for $0<\gamma<0.25$, $v_g > 1$, which is in contrast to $v_g =1$ in the case of systems with NN coupling; (ii) for $\gamma=0.25$, $v_g$ is very close to zero in a wider $k$ domain near the Brillouin zone boundary; (iii) for $\gamma>0.25$, both $\omega_q$ and $v_g$ are actually degenerate, i.e., one $\omega_q$/$v_g$ corresponds to two $q$ values. These unusual properties can favor the formation of a special highly moving localized excitation (mobile DBs) in the presence of appropriate nonlinearity\cite{Xiong-1,Xiong-2} and thus one can expect that such properties would greatly influence the energy relaxation.

\section{Method}
We use the cooling method to reveal the slow energy relaxation and to obtain the information of DBs. For this purpose, we first thermalize the lattices to a fixed temperature $T=0.1$ by using the Nose-Hoover\cite{Nose} heat baths. Then after the systems are fully thermalized for a considerably long time, we remove the heat baths and add the absorbing boundary conditions. That is to say, we impose damping to both ends (which are free) of the lattices, and so the equation of motion of the system can be expressed by
\begin{equation}
\ddot{x_i}= V'(x_{i+1}-x_i)+V'(x_{i}-x_{i-1})+\gamma V'(x_{i+2}-x_i)+\gamma V'(x_{i}-x_{i-2})- \eta p_i (\delta_{i,1}+\delta_{i,N}).
\end{equation}
Here $\ddot{x_i}=\frac{\mathrm{d} p_i(t)}{\mathrm{d} t}$; $V'(\xi)$ represents $\frac{\partial V}{ \partial \xi}$; $\eta$ is the dissipation exponent and we fix $\eta=0.1$ in practice; $\delta$ is the Kronecker delta.

By performing the cooling method, we are interested in the decay of the total energy. To reveal this decay, we first define the symmetrized site energies:
\begin{equation}
h_i=\frac{1}{2} p_i^2+\frac{1}{2} \left[ V(x_{i+1}-x_i)+V(x_{i}-x_{i-1})+\gamma V(x_{i+2}-x_i)+\gamma V(x_{i}-x_{i-2})\right ].
\end{equation}
With such definition, the total energy of the system is given by $E=\sum_{i=1}^N h_i$. We can now study the decay of the normalized energy $E(t)/E(0)$, which represents the ratio of the decayed energy to the initial energy. As it has been shown that in some cases the exponential law or the stretched-exponential law is observed. To identify what exactly the law is, it is better to plot
\begin{equation}
D(t)=-{\rm ln} \left[E(t)/E(0)\right].
\end{equation}

We remind that since the energy localization might be strongly inhibited by fixed-end boundary conditions\cite{Re-3,Re-4}, in our simulations we apply the free-end boundary conditions. The motion equations are integrated with the velocity-Verlet algorithm\cite{Verlet} with a time step $0.01$. With both heat baths and damping, the systems are first evolved $10^6$ times with the heat baths to reach the stationary state under the given temperature, and next evolved additional $10^6$ times without the heat baths but with the damping presented, which enables us to reveal the relaxation process induced by the absorbing boundaries.

\section{Results}
\subsection{Harmonic system}
\begin{figure}[bt]
\centerline{\epsfig{file=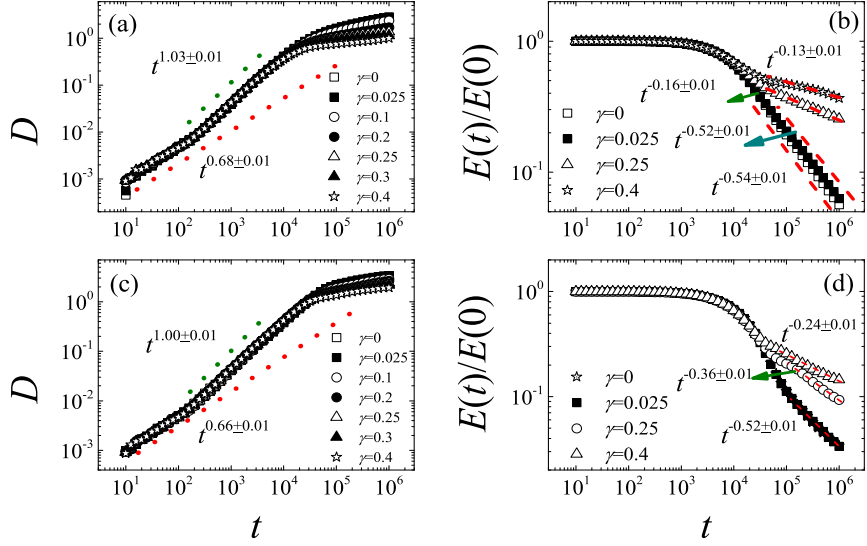,width=12cm,height=8cm,clip=}}
\vspace*{8pt}
\caption{(a) and (c): $D$ versus $t$ with several values of $\gamma$, where the two dashed lines denote the best fittings of $t^{\nu}$ and they indicate the previously predicted exponential law and the new observed stretched-exponential law decay of $E(t)/E(0)$, respectively. (b) and (d): $E(t)/E(0)$ versus $t$ under several $\gamma$, where the dashed lines denote the best fittings of $t^{-\mu}$ that indicates the power-law decay of $E(t)/E(0)$. Figures (a) and (b) are for the harmonic potential; Figures (c) and (d) are for the FPUT-$\beta$ potential.}
\label{fig2}
\end{figure}
To show the role of NNN coupling in slow energy relaxation separately, let us first check the results of the relevant linear systems with NNN coupling. For the same harmonic system with the NN coupling only, it has been theoretically predicted that \cite{Re-3}
\begin{equation}
\frac{E(t)}{E(0)}=
\begin{cases}
\mathrm{e}^{-t/\tau_0}&\mbox{for $t \ll \tau_0$},\\
[2 \pi (t/\tau_0)]^{-1/2}&\mbox{for $t \gg \tau_0$}.
\end{cases}
\end{equation}
This means that one generally finds two energy relaxation processes, i.e., the exponential law and the inverse-square-root law, for short and long timescales, respectively. Such two laws have been verified in a harmonic chain with NN coupling only under a small $N=32$, but some deviations have also been numerically observed\cite{Re-3,Re-4}, i.e., a new exponential relaxation recovers in a long time. In view of this, we study the dependence of the energy relaxation properties on $\gamma$ here in a large system size $N=4096$ in Fig.~\ref{fig2}(a). Surprisingly, in a time of $t <10^4$, regardless of $\gamma$, the results of $D$ versus $t$ follow two general power-laws ($\sim t^{\nu}$) with $\nu=0.68$ and $\nu=1.03$, respectively. This suggests that in addition to the previously theoretically predicted exponential law, there is another stretched-exponential law $\mathrm{e}^{-t^{\nu}/\tau_0}$ with $\nu=0.68$ at a shorter time, independent of the NNN coupling ratio $\gamma$.

Fig.~\ref{fig2}(a) also suggests that the long-time behavior of the energy relaxation should depend on $\gamma$. To see this more explicitly, Fig.~\ref{fig2}(b) depicts the results of $E(t)/E(0)$ versus $t$ in harmonic systems with NNN coupling. A log-log plot then helps us clearly identify several power-law exponents. The best fittings give $E(t)/E(0) \sim t^{-\mu}$ with $\mu=0.54$, $\mu=0.52$, $\mu=0.16$, and $\mu=0.13$ for $\gamma=0$, $\gamma=0.025$, $\gamma=0.25$, and $\gamma=0.4$, respectively (in a long time). It indicates that including the NNN coupling seems to further slow down the energy relaxation process. This is not trivial since from Fig.~\ref{fig1} by including the NNN coupling, on one hand the phonon group velocity $v_g$ is increased with the increase of $\gamma$. This speeds up the energy relaxation process. On the other hand, the increase of $\gamma$ causes more phonons with higher frequencies to emerge, which however slows down the relaxation. Therefore, in a linear system with NNN coupling but without nonlinearity and DBs, the overall effect is to depress the energy relaxation process during the cooling.

\subsection{FPUT-$\beta$ system}
With the above understanding, let us now consider the combined effects of NNN coupling and nonlinearity. For this purpose, as mentioned we focus on the popular 1D FPUT-$\beta$ lattice. Fig.~\ref{fig1}(c) shows that the first stretched-exponential and exponential laws of $E(t)/E(0)$ for $t < 10^4$, which are similar to that shown in Fig.~\ref{fig1}(a). Therefore, these two relaxation laws seem independent on the NNN coupling, even when one includes the nonlinearity into the systems. This may also suggest that such two laws  for $t < 10^4$ are general, at least for both the linear and nonlinear systems with NNN coupling. Therefore, the origin of this finding is worth studying in the future.

Again, the distinctions for different $\gamma$ in the FPUT-$\beta$ systems still lie in the long-time behavior. As can be seen in Fig.~\ref{fig2}(d), $E(t)/E(0)$ follows a $t^{-\mu}$ law at long timescales, similar to those observed in the harmonic systems as shown in Fig.~\ref{fig2}(b). A further comparison of Figs.~\ref{fig2}(b) and (d) shows that the power-law exponent $\mu$ for harmonic and FPUT-$\beta$ systems under the same $\gamma$ is different, i.e., $\mu$ for the FPUT-$\beta$ lattice is larger than that for harmonic lattice as long as $\gamma \neq 0$. It presents a clue that the combined effects of NNN coupling and nonlinearity seems to speed up the relaxation during the cooling.

\begin{figure}[bt]
\centerline{\epsfig{file=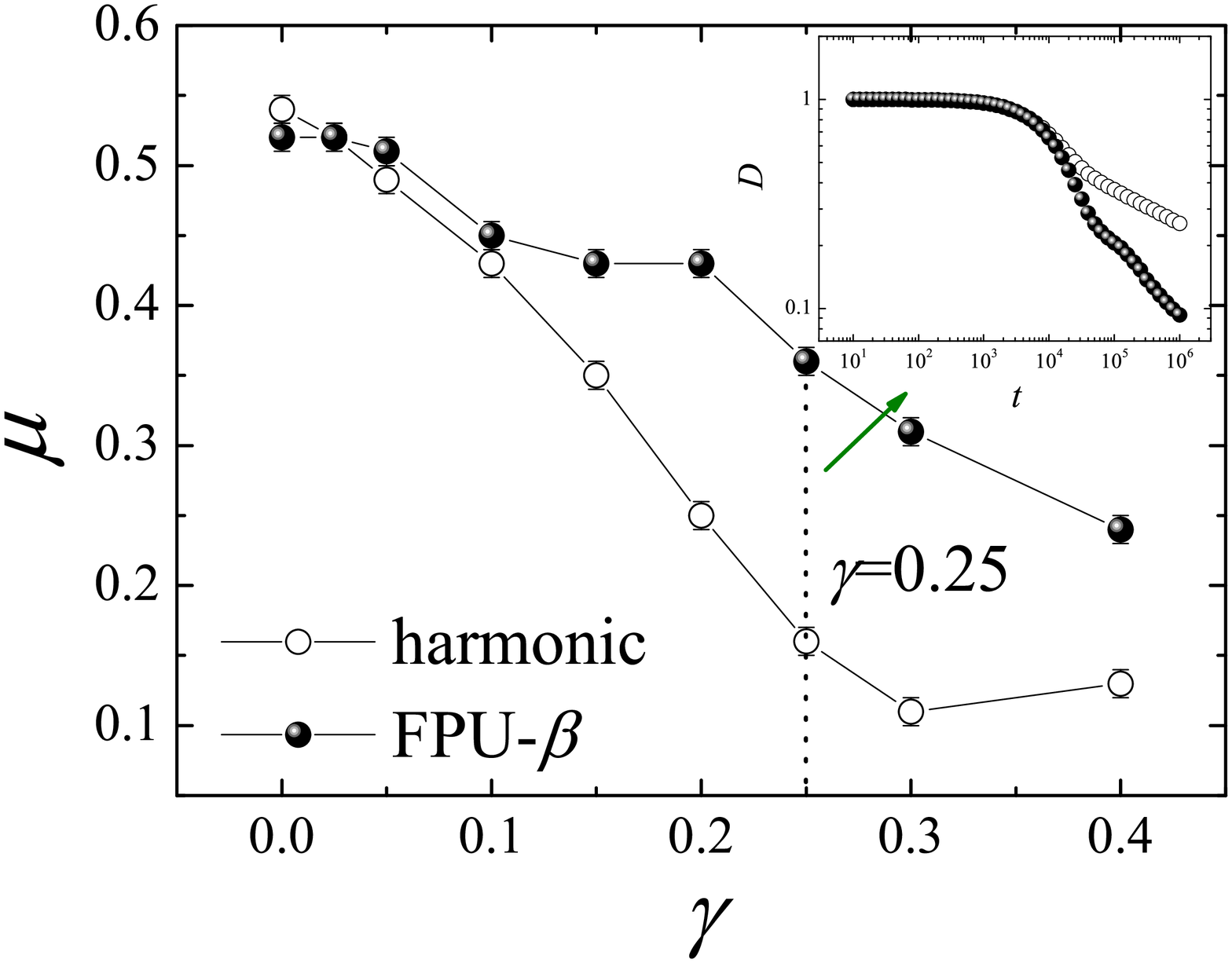,width=12cm,height=8cm,clip=}}
\vspace*{8pt}
\caption{$\mu$ versus $\gamma$, where the hollow and solid circles represent the results of harmonic and FPUT-$\beta$ systems, respectively; the dashed line indicates the results of $\gamma=0.25$. For a better illustration, we also plot $D$ versus $t$ for $\gamma=0.25$, separately in the inset.}
\label{fig3}
\end{figure}

In order to have a close look at this trend, in Fig.~\ref{fig3} we plot $\mu$ against $\gamma$ for both the harmonic and FPUT-$\beta$ lattices. Therein four data points are extracted from Figs.~\ref{fig2}(b) and~\ref{fig2}(d), while others are calculated additionally in the same way. Indeed, in the range of $\gamma$ investigated, all $\mu$ of FPUT-$\beta$ systems are larger than those of the harmonic systems. Indeed, this indicates that the combined effects of the NNN coupling and nonlinearity are to make the energy relaxation relatively faster if compared with the counterpart linear systems.

\subsection{The origin of the relatively faster energy relaxation: the role of mobile DBs}

\begin{figure}[bt]
\centerline{\epsfig{file=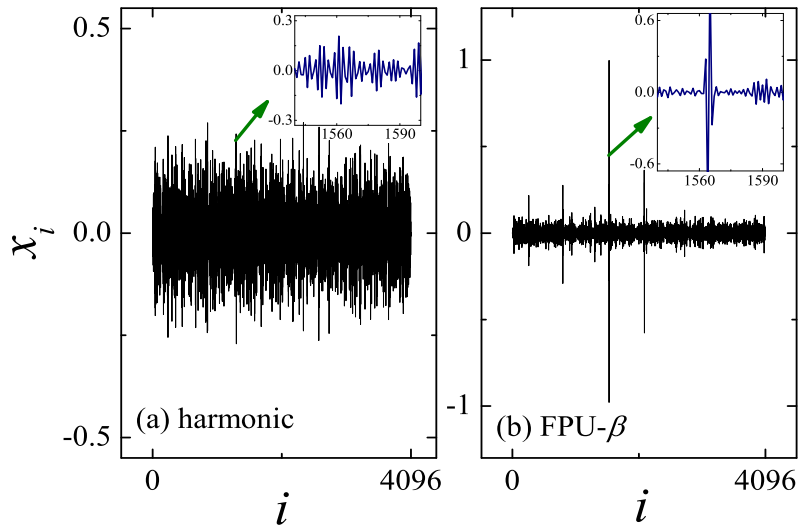,width=12cm,height=8cm,clip=}}
\vspace*{8pt}
\caption{Snapshot of atomic displacement after imposing the damping to the free-end boundaries by a time $t=10^{6}$, where left(right) panel corresponds to a harmonic (FPUT-$\beta$) system with NNN coupling ($\gamma=0.25$). The inset in each panel
is a zoom for the reference sites.}
\label{fig4}
\end{figure}

We lastly explain the observed relatively faster energy relaxation in a nonlinear system compared to the counterpart linear systems. As it has already been pointed out in introduction, the slow energy relaxation behaviors in nonlinear lattices are usually related to the energy localization and this localization origins from the excitation of standing DBs. In the absence of nonlinearity, it was previously regarded that the energy relaxation will turn back to the traditional exponential decay~\cite{Re-3}. However, our above results of Fig.~\ref{fig2} suggest that even for linear systems, this is not the case. The long-time behavior of the energy relaxation seems always to be a power-law decay. Combining the results of linear and nonlinear systems (see Figs.~\ref{fig2}(a)-(d)), one might conclude that the power-law energy relaxation in a long time is general but the details of the power-law exponent are then affected by the localization induced by DBs. To illustrate this point, in Fig.~\ref{fig4} we plot a snapshot of the lattice displacements at the end of the simulation, i.e., by imposing the damping to both harmonic and FPUT-$\beta$ systems after a time $10^6$ (with $\gamma=0.25$ for an example). As can be seen, only in the FPUT-$\beta$ system we identify the emergence of DBs [see the inset of Fig.~\ref{fig4}(b), a profile of DBs can be found there], which is in clear contrast to that observed in the harmonic system [see the inset of Fig.~\ref{fig4}(a)].

\begin{figure}[bt]
\centerline{\epsfig{file=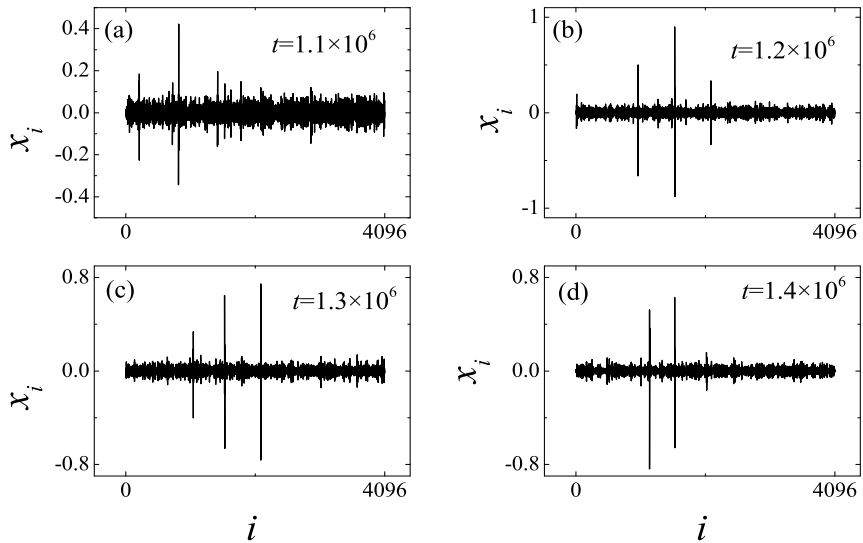,width=12cm,height=8cm,clip=}}
\vspace*{8pt}
\caption{Snapshot of atomic displacement after imposing the damping to the free-end boundaries by a time $t=1.1 \times 10^{6}$ (a), $t=1.2 \times 10^{6}$ (b), $t=1.3 \times 10^{6}$ (c), and $t=1.4 \times 10^{6}$ (d) in the FPUT-$\beta$ system with NNN coupling ($\gamma=0.25$).}
\label{fig5}
\end{figure}

Since DBs can localize energies, one may wonder why the energy relaxation in the nonlinear systems with NNN coupling is relatively faster than that in the counterpart linear systems. This can be understood from Fig.~\ref{fig5} if we further examine the mobility of DBs. In Fig.~\ref{fig5} we plot the same snapshot of the lattice displacements after imposing the damping the free-end boundaries of the nonlinear lattice ($\gamma=0.25$) for four long times. It can be seen that due to the mobility of DBs, they can interact with both phonons and boundaries, which produces an additional passageway for speeding up the energy relaxation. This is surely different from the main passageway of phonons-boundaries scattering as shown in the relevant linear model (see Fig.~\ref{fig4}(a)).
\section{Conclusion}
To summarize, by applying a damping term into the system's boundaries, we have numerically studied the energy relaxation processes in both 1D linear and FPUT-$\beta$ lattices when the NNN interactions are considered. As we expect that including the NNN coupling might need a longer time for systems to relax,
we have considered a system size much larger than that was taken account of in previous studies. For both linear and nonlinear lattices, generally the short-time ($t < 10^4$) relaxation behaviors are given by a stretched-exponential law, followed by an exponential law, which are independent of the ratio of the NNN coupling. This justifies that even linear systems can also support the stretched-exponential relaxation law at some timescales. Furthermore, the long-time ($t > 10^4$) behavior is however determined by a power-law with the exponent strongly dependent on $\gamma$, i.e., as $\gamma$ increases, the relaxation is slowed down. This means that including the NNN coupling makes the long-time power-law energy relaxation slower. Further introducing the nonlinearity, however, in turn, speed up the relaxation, which is induced by the emergence of mobile DBs. These mobile DBs can interact both phonons and boundaries, which then produce an additional passageway for scattering the energies. Finally, this novel energy relaxation behavior, induced by mobile DBs in the nonlinear lattices with NNN coupling, are surely different from the stretched-exponential relaxation observed in lattices with on-site potentials that is mainly produced by the standing DBs\cite{Re-1,Re-2}.





\begin{thebibliography}{99}
\vspace*{-1mm}
\begin{small}\baselineskip=10pt\itemsep-2pt
\bibitem{DB-1} A. J. Sievers and S. Takeno, {\it Phys. Rev. Lett.} {\bf 1}, 970 (1988).

\bibitem{DB-2} J. B. Page, {\it Phys. Rev. B} {\bf 41}, 7835 (1990).

\bibitem{DB-3} S. Flach and C. R. Willis, {\it Phys. Rep.} {\bf 295}, 181 (1998).

\bibitem{DB-4} S. Flach and A. V. Gorbach, {\it Phys. Rep.} {\bf 467}, 1 (2008).

\bibitem{DB-5} M. Sato, B. E. Hubbard, and A. J. Sievers,  {\it Rev. Mod. Phys.} {\bf 78}, 137 (2006).

\bibitem{DB-6} E. Trias, J. J. Mazo, and T. P. Orlando, {\it Phys. Rev. Lett.} {\bf 84}, 741 (2000).

\bibitem{DB-7} P. Binder, D. Abraimov, A. V. Ustinov, S. Flach, and Y.
Zolotaryuk, {\it Phys. Rev. Lett.} {\bf 84}, 745 (2000).

\bibitem{DB-8} F. Palmero, L. Q. English, X.-L. Chen, W. Li, J. Cuevas Maraver, and P. G. Kevrekidis, {\it Phys. Rev. E} {\bf 99}, 032206 (2019).

\bibitem{DB-9} A. Gomez-Rojas and P. Halevi, {\it Phys. Rev. E} {\bf 97}, 022225 (2018).

\bibitem{DB-10} Y. Watanabe, T. Nishida, Y. Doi, and N. Sugimoto, {\it Phys. Lett. A} {\bf 382}, 1957 (2018).


\bibitem{DB-11} J. Cuevas, L. Q. English, P. G. Kevrekidis, and M. Anderson, {\it Phys. Rev. Lett.} {\bf 102}, 224101 (2009).

\bibitem{DB-12} F. M. Russell, Y. Zolotaryuk, J. C. Eilbeck, and T. Dauxois, {\it Phys. Rev. B} {\bf 55}, 6304 (1997).

\bibitem{DB-13} K. Vorotnikov, Y. Starosvetsky, G. Theocharis, and P. G.
Kevrekidis, {\it Physica D} {\bf 365}, 27 (2018).


\bibitem{DB-14} C. Chong, M. A. Porter, P. G. Kevrekidis, and C. Daraio,
{\it J. Phys.: Condens. Matter} {\it 29}, 413003 (2017).

\bibitem{DB-R} S. V. Dmitriev, E. A. Korznikova, J. A. Baimova, and M. G.
Velarde, {\it Phys. Usp.} {\bf 59}, 446 (2016).

\bibitem{DB-15} L. Z. Khadeeva and S. V. Dmitriev, {\it Phys. Rev. B} {\bf 81}, 214306 (2010).

\bibitem{DB-16} A. Riviere, S. Lepri, D. Colognesi, and F. Piazza, {\it Phys. Rev. B} {\bf 99}, 024307 (2019).


\bibitem{DB-17} R. T. Murzaev, D. V. Bachurin, E. A. Korznikova, and S. V.
Dmitriev, {\it Phys. Lett. A} {\bf 381}, 1003 (2017).

\bibitem{DB-18} R. T. Murzaev, R. I. Babicheva, K. Zhou, E. A. Korznikova,
S. Y. Fomin, V. I. Dubinko, and S. V. Dmitriev, {\it Eur. Phys. J. B} {\bf 89}, 168 (2016).

\bibitem{DB-19} V. Dubinko, D. Laptev, D. Terentyev, S. V. Dmitriev, and
K. Irwin, {\it Comp. Mater. Sci.} {\bf 158}, 389 (2019).

\bibitem{DB-20} B. Juanico, Y.-H. Sanejouand, F. Piazza, and P. De Los Rios,
Discrete Breathers in Nonlinear Network Models of Proteins,
{\it Phys. Rev. Lett.} {\bf 99}, 238104 (2007).

\bibitem{DB-21} C. L. Gninzanlong, F. T. Ndjomatchoua, and C. Tchawoua,
Forward and backward propagating breathers in a DNA model
with dipole-dipole long-range interactions, {\it Phys. Rev. E} {\bf 102},
052212 (2020).

\bibitem{Re-1} G. P. Tsironis and S. Aubry, {\it Phys. Rev. Lett.} {\bf 77}, 5225 (1996).

\bibitem{Re-2} A. Bikaki, N. K.  Voulgarakis, S. Aubry, and G. P. Tsironis, {\it Phys. Rev. E} {\bf 59}, 1234 (1999).

\bibitem{Re-3} F. Piazza, S. Lepri, and R. Livi, {\it J. Phys. A} {\bf 34}, 9803 (2001).

\bibitem{Re-4} F. Piazza, S. Lepri, R. Livi, {\it Chaos} {\bf 13}, 637 (2003).

\bibitem{Xiong-1} D. Xiong, J. Wang, Y. Zhang, and H. Zhao, {\it Phys. Rev. E} {\bf 85}, 020102 (2012).

\bibitem{Xiong-2} D. Xiong, Y. Zhang, and H. Zhao, {\it Phys. Rev. E} {\bf 90}, 022117 (2014).

\bibitem{Xiong-3} D. Xiong, Y. Zhang, and H. Zhao, {\it Phys. Rev. E} {\bf 88}, 052128 (2013).

\bibitem{Onsite-1} C. Giardin\`{a} R. Livi, A. Politi, M. Vassalli, {\it Phys. Rev. Lett.} {\bf 84}, 2144 (2000).

\bibitem{Onsite-2} O.V. Gendelman, A.V. Savin, {\it Phys. Rev. Lett.} {\bf 84} 2381 (2000).

\bibitem{Onsite-3} D. Xiong, D. Saadatmand, and S. V. Dmitriev, {\it Phys. Rev. E} {\bf 96} 042109 (2017).

\bibitem{LR} J. Wang, S. V. Dmitriev, and D. Xiong {\it Phys. Rev. Res.} {\bf 2} 013179 (2020).

\bibitem{Scatter} D. Saadatmand, D. Xiong, V. A. Kuzkin, A. M. Krivtsov, A. V.
Savin, and S. V. Dmitriev, {\it Phys. Rev. E} {\bf 97}, 022217
(2018).

\bibitem{Others-1} M. E. Manley, {\it Acta Mater.} {\bf 58}, 2926 (2010).

\bibitem{Others-2} E. A. Korznikova, A. Y. Morkina, M. Singh, A. M. Krivtsov,
V. A. Kuzkin, V. A. Gani, Yu. V. Bebikhov, and S. V. Dmitriev, {\it Eur. Phys. J. B} {\bf 93}, 123 (2020).

\bibitem{Others-3} M. E. Manley, M. Yethiraj, H. Sinn, H. M. Volz, A. Alatas,
J. C. Lashley, W. L. Hults, G. H. Lander, D. J. Thoma, and J. L.
Smith, {\it J. Alloy. Compd.} {\bf 444}, 129 (2007).

\bibitem{Others-4} B. Mihaila, C. P. Opeil, F. R. Drymiotis, J. L. Smith, J. C.
Cooley, M. E. Manley, A. Migliori, C. Mielke, T. Lookman,
A. Saxena et al., {\it Phys. Rev. Lett.} {\bf 96}, 076401 (2006).

\bibitem{Others-5} M. Singh, A. Y. Morkina, E. A. Korznikova, V. I. Dubinko, D.
A. Terentiev, D. Xiong, O. B. Naimark, V. A. Gani, and S. V.
Dmitriev, {\it J. Nonlinear Sci.} {\bf 31}, 12 (2021).

\bibitem{Nose} S. Nose, {\it J. Chem. Phys.} {\bf 81}, 511 (1984); W. G. Hoover,
{\it Phys. Rev. A} {\bf 31}, 1695 (1985).

\bibitem{Verlet} P. Allen and D. L. Tildesley, Computer Simulation of Liquids Clarendon, Oxford, 1987.
\end{small}
\end{thebibliography}
\end{document}